# High-speed Opto-electronic Pre-processing of Polar Mellin Transform for Shift, Scale and Rotation Invariant Image Recognition at Record-Breaking Speeds

**Julian Gamboa,**
**Xi Shen, Tabassom Hamidfar, Selim M. Shahriar**
*Northwestern University*

**ABSTRACT**

Space situational awareness demands efficient monitoring of terrestrial sites and celestial bodies, necessitating advanced target recognition systems. Current target recognition systems exhibit limited operational speed due to challenges in handling substantial image data. While machine learning has improved this scenario, high-resolution images remain a concern. Optical correlators, relying on analog processes, provide a potential alternative but are hindered by material limitations. Recent advancements in hybrid opto-electronic correlators (HOC) have addressed such limitations, additionally achieving shift, scale, and rotation invariant (SSRI) target recognition through use of the polar Mellin transform (PMT). However, there are currently no techniques for obtaining the PMT at speeds fast enough to take advantage of the inherent speed of the HOC. To that end, we demonstrate an opto-electronic PMT pre-processor that can operate at record-breaking millisecond frame rates using commercially available components for use in an automated SSRI HOC image recognition system for space situational awareness.

## 1. INTRODUCTION

Space situational awareness requires the monitoring of numerous terrestrial sites and spatial bodies. However, current state-of-the-art target recognition systems are severely limited in their operational speed, largely due to technological limitations when dealing with the large amounts of data inherent to image processing. New tools such as machine learning have undoubtedly helped alleviate these problems, but themselves face speed issues when using images with higher resolutions. The current leading computational algorithm based on machine learning, YOLOv7, is capable of achieving a 51.4% object detection rate at 118 frames per second (fps) at an image size of 640 pixels [1]. However, when the resolution is increased to 1280 pixels, the speed drops drastically to 80 fps. In contrast, optical systems are largely independent of the resolution of the source images, as they primarily rely on analog physical processes for operation. Optical correlators, for example, use a combination of lenses and holograms to perform Fourier transforms (FTs) and optically multiply them together, producing a correlation signal. This technology has existed for many decades, yet it has remained impractical due to performance limitations of the required dynamic nonlinear materials [2].

Research into high-speed optical correlators has primarily focused on finding a way to overcome these material limitations. To that end, in recent years we have proposed [3] and demonstrated [4] a hybrid opto-electronic correlator (HOC) that achieves this by using spatial light modulators (SLMs), phase stabilization circuits, and electronics to record the required amplitude and phase information of the FTs of input images on focal plane arrays (FPAs) through interference with reference plane-waves, thus replacing the analog holographic technique with a digital one. This correlator can theoretically operate at speeds on the microsecond scale but is limited by the availability of high-speed SLMs to project the query images into the optical domain. As such, holographic memory devices (HMDs) can be employed as all-optical databases of reference images within the HOC, eliminating the need for ultrafast SLMs [5]. Additionally, we have recently eliminated the requirement for optical phase stabilization in this architecture through the use of an off-axis technique, removing perhaps the largest practical limitation of this technology [6].

Because of the properties of the FT, optical correlators are inherently shift-invariant. However, any rotation of an input image will result in a directly proportional rotation of the FT, preventing otherwise identical images from matching. Similarly, if an input image is scaled, the FT will undergo an inversely proportional descaling, also preventing a match. In order to achieve shift, scale, and rotation invariant (SSRI) correlation, it is thus necessary to pre-process the input images to convert any rotation and scaling into linear shifts. The polar Mellin transform (PMT) is an especially good candidate for this pre-processing as it can be calculated as the log-polar transform (LPT) of the magnitude of the FT of an image [7,8]. Because of this, optics can be used for the FT part of the operation, while an electronic module can be used for the log-polar component. This PMT pre-processing can be performed ahead of time for the reference arm of the HOC, where the pre-processed images can be stored directly on an HMD to later be accessed at high speed. This is not possible for the query arm however, as the query images are by definition not a-priori known. For this reason, it is necessary to develop an opto-electronic PMT pre-processor (OPP) that can operate at speeds compatible with the rest of the HOC.

The rest of the paper is structured as follows. Section 2 gives an overview of the PMT and opto-electronic SSRI target recognition. The design of the OPP using commercially available components is covered in section 3. Experimental results showing real-world PMT conversion times below 2 ms are described in section 4. Finally, conclusions and outlook are given in section 5.

## 2. SSRI TARGET RECOGNITION VIA THE POLAR MELLIN TRANSFORM

The theory and architecture of the HOC is described in more detail in references [3,6,9]. A diagram of this architecture is shown in Fig. 1. Briefly, the HOC functions as follows. First, a laser beam is expanded and split into an input arm, an output arm, and an OPP arm. The input arm consists of two identical segments, one for the reference and another for the query, wherein an HMD or SLM projects an image towards a lens that produces the FT at its output plane. One FPA detects the intensity of the FT, another detects the intensity of the interference between the FT and an auxiliary plane-wave, and a third FPA detects the intensity of the auxiliary plane-wave itself. The detected signals are processed electronically and projected on an output SLM to be optically FT'd again, producing the final correlation signal that can be captured on an FPA or a detector, depending on the application. In order to avoid having to stabilize the optical phases of the auxiliary beams, it is necessary to place the reference and query HMD/SLM in a manner such that they are shifted in opposite directions along their projection planes. The OPP arm is identical in construction to the output arm, consisting of an SLM, a lens, and an FPA. In this case, however, the SLM projects a raw query image from an outside source such as a camera or a database, and an electronic module receives the signal from the FPA that contains the intensity of the FT of the query. The module then processes this information via the LPT to yield the PMT, as described in more detail below, sending the resulting image to the input SLM of the query arm.

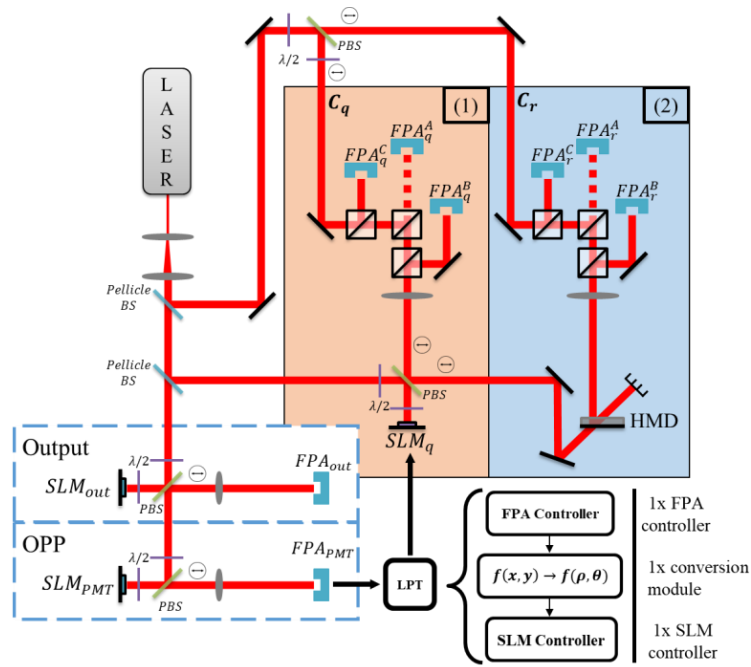

Fig. 1. Architecture of the phase insensitive HOC. (1): Query image arm using an SLM as an input source. The SLM is connected to an electronic module that provides the PMT of an image input at the OPP arm. (2): Reference image arm using an HMD as an input source.

In general, the PMT can be obtained through the following process:
1) Calculate the FT of the original image, $G$.
2) Obtain the magnitude of the FT.
3) Define (x=0, y=0) for the FT plane to be located at the center of the FT.
4) Calculate the LPT of the magnitude of the FT as follows:
    a. The horizontal axis of the LPT represents $\rho$, the logarithm of the radius ($r$) of each pixel in the FT:
    $$\rho = \ln\left(\frac{r}{r_0}\right) = \ln\left(\frac{1}{r_0}\sqrt{x^2 + y^2}\right).$$
    where $r_0$ is the minimum radius to be transformed. The horizontal axis ranges from $\rho = 0$ (such that $r = r_0$) to that obtained at the maximum values of x and y ($r = r_{max}$). Circular DC blocking is also necessary, as the DC component of the FT would otherwise dominate the PMT. Thus, the calculation is only performed for $r \geq r_{dc}$, where $r_{dc}$ is selected to block out the DC component without eliminating additional information, and $r_{dc} \geq r_0$.
    b. The vertical axis represents $\theta$, the angle of each pixel in the FT, and ranges from 0 to $2\pi$:
    $$\theta = \arctan(y/x)$$

If an image is scaled up by a multiplicative factor of '$a$', its FT will be scaled down by the same factor. Correspondingly, the value of $x$ & $y$, and thus $r$, for the scaled FT will be divided by '$a$'. Carrying this through to the definition of $\rho$, we see that the scaling factor is effectively converted into a horizontal linear shift: $\rho(r/a) = \rho(r) - \ln(a)$. Similarly, if an image is rotated by a factor '$\phi$', its FT will experience an identical rotation. Thus, the $\theta$ axis will be linearly shifted by this value, bearing in mind that the axis is modulo $2\pi$. Therefore, the PMT of an image converts any rotation and scaling into orthogonal linear shifts, which correspondingly shift the cross-correlation and convolution at the output plane. As a tradeoff, the position of the target in the original image is not carried over when performing the PMT due to the fact that this information is contained within the FT by means of the phase, which is eliminated when measuring its intensity. As a result, the position of the correlation output can be used to directly identify any rotation and scaling between two matched inputs, but the location of the target can no longer be determined with this same information. This compromise between being able to detect a target in an SSRI manner and being able to identify its position on the input plane can be overcome by using the information obtained about the differences in scale and rotation between the query and the reference images to rescale and re-orient the query image such that it matches the scale and rotation of the reference image. Once this is done, one can then use this query image (without PMT processing) as the input to the HOC, thus enabling the determination of any shift of the query image. Alternatively, it is possible to combine the HOC with a more general image recognition system, such as YOLOv7. In this scheme, the optical correlator is used as an input filter which can select images that merit further processing from a large query dataset, and the machine learning-based processor follows up to extract more precise information about the target.

## 3. OPTO-ELECTRONIC PMT PRE-PROCESSOR

In the previous section, we discussed how the PMT allows a correlator to perform SSRI target recognition, whereby the original image must be transformed prior to being correlated. In principle, this could be performed computationally, but would incur significant processing delays due to the use of digital approximations of the FT. Clearly, the relationship between the PMT and the FT makes it especially well-suited for optical techniques. Nonetheless, the log-polar operation that converts the FT into the PMT is not realizable optically, and so requires additional electronic processing to complete. Fig. 2 shows a diagram of the OPP architecture reported here, wherein the FT is obtained optically via a lens, detected opto-electronically via a high-speed FPA, processed electronically, and finally projected opto-mechanically via a digital micromirror device (DMD) SLM.

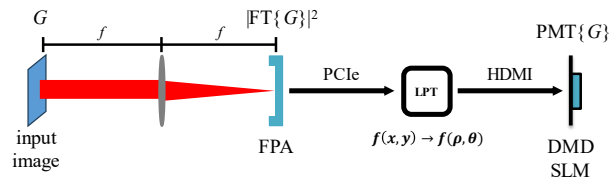

Fig. 2. Opto-electronic PMT Pre-processor (OPP) architecture. The original image, G, is optically input to the OPP one focal distance, $f$, away from a convex lens using a collimated laser source. The intensity of the FT is detected by a high-speed FPA that is placed on the opposite focal plane of the lens. An electronic module performs the LPT of the detected Fourier spectrum to obtain the PMT of G. The result is projected back into the optical domain via a high-speed digital micromirror device (DMD) SLM.

In order for this architecture to function at high-speed, the choice of components is critical. Currently, FPAs that can capture thousands of frames per second at Full HD (1920x1080) resolution are commercially available, and typically interface with electronic hardware via the standard PCI-express (PCIe) interface. Unfortunately, SLMs of similar speeds are still under development and so present a significant bottleneck. DMD SLMs are among the fastest currently available on the market, limited primarily by the transmission data-rate of the selected input standard. For the experiments presented here, a Texas Instruments DLP471TE SLM was selected, which uses HDMI 2.0 and can reach up to 240 fps for full-color (RGB888) frames at Full HD. Because of the design of DMD SLMs, a full-color frame is displayed one color at a time, first projecting the red channel, then the green channel, and finally the blue channel. This allows us to use one full-color frame to project three monochrome frames, effectively tripling the maximum framerate to 720 fps while still taking advantage of the HDMI standard. For the work presented here, the SLM was operated at 166-hz, full-color, and Full-HD, achieving a true framerate of 498 fps.

The high-speed FPA that was selected uses PCIe direct memory access (DMA) to allow the processing module to read the captured frames from its on-board random access memory (RAM) to a local memory module without requiring additional processing. While otherwise beneficial, this technique complicates the use of a field programable gate array (FPGA) as the processing module and instead incentivizes the use of a CPU for the same task. While a CPU is convenient for its ease of programming, an FPGA has a faster potential output due to its capacity for parallel processing, and an application specific integrated circuit (ASIC) would provide the fastest possible speed at the cost of custom semiconductor manufacturing. However, the pursuit of the use of the FPGA or ASIC approaches have to wait until ultra-high-speed SLMs become available in the future.

### 3.1 CPU-BASED OPP DESIGN

The LPT is a projection of a polar coordinate system on a cartesian coordinate system, where the radius and angle are expressed as perpendicular axes, and the former is on a logarithmic scale. Moreover, this is exclusively a coordinate transformation, and so the relationship between the input and output of an LPT is independent of the particular value of any given pixel. Thus, for a given output pixel located at a coordinate ($\rho,\theta$), the value will always be given by the same input pixel at a coordinate ($x,y$), regardless of the value itself. Because of this property, the precise input-output relationship can be pre-calculated for each pixel and programmed directly into the LPT module, such that it is not necessary to perform any calculations during its operation. Ideally, this would allow us to construct an ASIC that physically wires the input pixel registers to their corresponding output pixel registers, allowing for the LPT to be performed in a single clock cycle. An approximation of this design can be constructed on an FPGA, allowing for hundreds of pixels to be transformed at each clock cycle, but requires a barebones FPA that would be tightly integrated into the hardware design. Alternatively, a CPU-based approach can be constructed to take advantage of the PCIe-based DMA that is available on commercial high-speed FPAs, while simultaneously allowing the OPP to be integrated easily into existing image processing ecosystems. Moreover, DMA does not use any CPU resources, requiring only an initialization to allow the operating system to access the FPA memory as if it were local memory. Additionally, modern CPUs are constructed with multiple cores that can run in parallel, allowing us to design a moderately parallelized architecture that is fast enough to reach the limits of modern SLMs.

The SLM selected for these experiments was operated at 166 Hz, full color, and Full HD, allowing for the display of up to 498 frames of 1920 x 1080 pixels each second, where the amplitude of each pixel is represented in 8 bits. This is equivalent to a frame time of just over 2 ms, which represents the maximum amount of time the program can take to capture a FT, process it, and display the LPT. As mentioned earlier, the SLM can only receive an RGB frame at 166 fps, where each color channel will be displayed sequentially, hence tripling the effective framerate. For this reason, it is necessary to capture and process three independent monochrome frames that will be combined into a single RGB frame that can be transmitted via HDMI. Thus, it is more accurate to list the RGB frame time of 6 ms as the maximum amount of time to capture three FTs, process them, combine the LPTs into a single frame, and display the results. Fig. 3 shows a general flowchart for this process, labelling the execution times that are critical to achieving the desired framerate. The time it takes to capture a FT frame, labelled $t_{cap}$, will depend on the configured exposure time, readout time, and transfer time from the FPA RAM to the local RAM. Typically, the exposure time will be much lower than the readout and transfer times and can be further tuned by adjusting the input laser power. The LPT processing time, labelled as $t_{lpt}$, presents the primary challenge of using a CPU-based approach, as this step can take a significant amount of processing. Finally, the image display time, labelled $t_{dis}$, will depend primarily on the selected framerate.

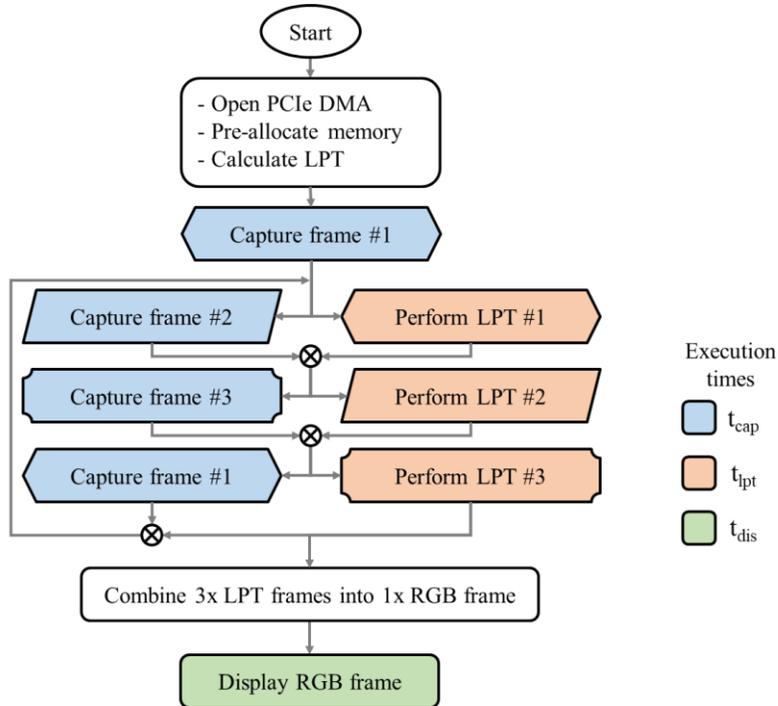

Fig. 3. Flowchart of the CPU-based OPP logic. The colored blocks represent segments that can take a significant amount of processing time. The execution times are per-block. Here, the ⊗ symbol represents a "wait until" operation, which does not continue until both of the input blocks are completed. A block with two outputs represents a parallel execution of the two following blocks. Red blocks with a similar shape to a blue block receive data from the latter.

The LPT architecture described here is constructed to take advantage of the fact that the input-output coordinate relationship is a priori known and can be stored in local memory to be accessed at high-speed. It is convenient to store this data such that the output coordinates ($\rho, \theta$) can be used to find the matching input coordinates ($x,y$). This can be accomplished by using a 3D array, labelled *map*, of size [$\rho_{size},\theta_{size},2$], where ($\rho,\theta,1$) and ($\rho,\theta,2$) contain the corresponding $x$ & $y$ input coordinates, respectively. The LPT can then be performed by scanning each of the output coordinates ($\rho, \theta$) and performing the remapping according to the relationship stored in the *map* array:

$$Output(\rho,\theta) = Input(x = map(\rho,\theta,1), \quad y = map(\rho,\theta,2))$$

This approach allows us to construct the LPT function using memory-based programming, which can be significantly faster than object-based programming. For example, the C and C++ programming languages could perform this function using pointers and a for-loop. Additionally, the fact that the output only depends on a simple memory mapping means that the program can take advantage of CPU multithreading to drastically reduce the execution time. Additionally, because the system requires circular DC blocking, the $\rho$ coordinate only needs to be scanned for values of $\rho(r) \geq \rho(r_{dc})$, reducing the total number of operations when a larger $r_{dc}$ is selected.

The LPT module can be further optimized by running multiple processes in parallel. For example, the frame capture and PCIe DMA transfer do not utilize CPU resources, unlike the LPT function which relies exclusively on the CPU and memory, allowing both tasks to be performed simultaneously. It should be noted, however, that a specific frame cannot be captured and LPT'd at the same time, but rather that the LPT must be operating on a previously acquired frame, leading to the staggered architecture shown in Fig. 3. Here, the n-th frame is captured while the (n-1)-th frame is being LPT'd. Once both tasks are finished, the system then proceeds to capture the (n+1)-th frame and transform the n-th frame. When three frames have been captured and transformed, they are combined into a single RGB frame and transmitted to the DMD SLM, which then displays each color channel sequentially. This approach effectively means that only the larger of $t_{cap}$ and $t_{lpt}$ has any effect on the total execution time of the OPP. Furthermore, the SLM display is performed while the next set of three frames is being captured, setting the maximum execution time of the 3-frame loop to be equal to $t_{dis}$ at 6ms.

## 4. EXPERIMENTAL RESULTS

The CPU-based OPP design described in the previous section was constructed and tested. Fig. 4 shows the result of using the CPU-based electronic module without an optical input to process the LPT of three sample images. The images were selected to best exemplify the properties of the LPT and represent what would typically be the intensity of the FT of a real image. In these results, the effect of scaling and rotation are clearly visible, where a change in scale results in a linear horizontal shift, and a change in angle results in a vertical shift (modulo $2\pi$). This is also clearly visible in the combined RGB frame that was generated as a combination of the three monochrome LPT frames.

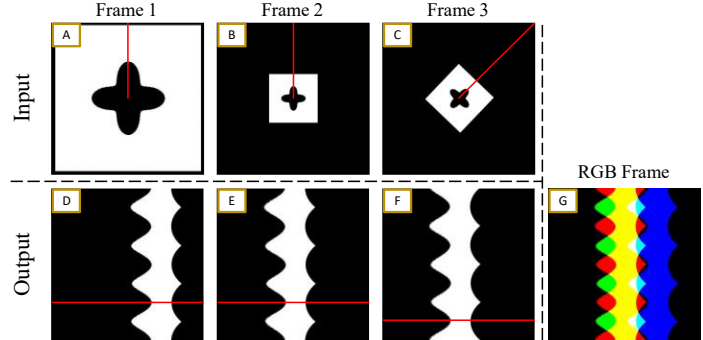

Fig. 4. Electronically processed LPT with the CPU-based electronic module of the OPP. (A,B,C): Sample images input into the CPU. The red lines mark the 'top' of the shape. (D,E,F): Output LPTs corresponding to (A), (B), and (C), respectively, where the images are in the ($\rho,\theta$) coordinate system. The red lines mark the same spot as their respective input images, showing the effect of rotation on the LPT. (G): RGB frame resulting from the combination of (D) in the blue channel, (E) in the green channel, and (F) in the red channel.

After validating the functionality of the electronic module, the full CPU-based OPP was constructed, using an additional SLM to optically input the original images into the system. The system used a Ryzen 9 3900x (12-core, 24-thread) CPU running Windows 10. The input SLM cycled through various real-world images whose FTs were captured on a high-speed FPA. The electronic module then accessed the measured FTs via PCIe DMA and performed the LPT as described in the previous section, using the scheme presented in Fig. 3. The result was output through a high-speed DMD SLM three frames at a time at a repetition rate of 166 RGB-frames per second, corresponding to 498 monochrome LPT-frames per second.

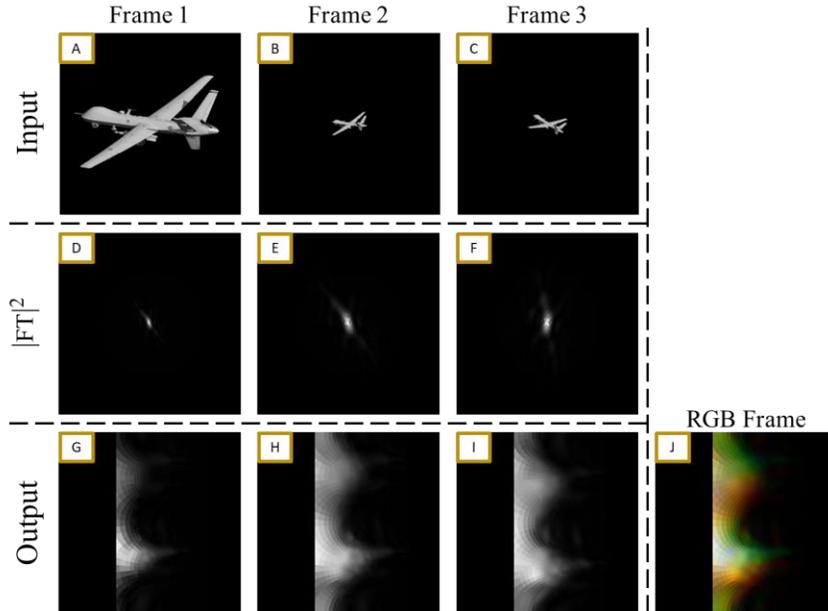

Fig. 5. Examples of the output from the CPU-based OPP. (A,B,C): Original image optically input into the OPP using an SLM. (D,E,F): Measured intensity of the FTs of the original images. (G,H,I): Result of the LPT, corresponding to the PMT of the original images. (J): RGB frame resulting from the combination of (G) in the blue channel, (H) in the green channel, and (I) in the red channel. Circular DC blocking was performed at a radius of 4 pixels for all cases.

Fig. 5 shows an example of the PMTs that were obtained through this OPP. Here, three real-world images of a Reaper drone were projected at different scales and angles. Similarly to Fig. 4, the effects of scaling and rotation are clearly visible in the output PMT. The system was operated over a period of a few minutes to measure the real-world value of $t_{cap}$ and $t_{lpt}$. From over 150,000 frames, the average DMA readout time was measured to be ~701 µs, and the average LPT conversion time was measured to be 1.636 ms. The latter value was later reduced to 983 µs by increasing the DC-blocking radius, $r_{dc}$, from 2 pixels to 4 pixels. These results show that even without the capture-plus-LPT parallelization described in Fig. 3, the total time between capturing a frame and processing the LPT is well below 2ms, successfully achieving a total PMT conversion framerate of 498 fps.

## 5. CONCLUSIONS AND OUTLOOK

A CPU-based OPP that can operate at millisecond speeds was designed and constructed using commercially available components, enabling real-time conversion of images into PMT signatures that can be used for opto-electronic SSRI target recognition. Demonstration of the operation of the OPP at this speed establishes the feasibility of operating an SSRI HOC essentially at the same speed, using only components that are currently available on the market. The 498 fps operating speed for images with a resolution of 1920 x 1080 represents nearly an order of magnitude enhancement in the speed of operation for the task of SSRI image recognition when compared to the leading computational methods. This speed can be enhanced further to near 1000 fps, being limited primarily by the availability of similarly fast SLMs. To date, this is the only optics-based PMT processor ever developed, enabling the fastest SSRI implementation in a correlator. In addition to enhancing the HOC, the OPP can be easily implemented in existing optical image processors to enable them to perform at high speed in an SSRI manner. In principle, the speed could be improved to operate in the microsecond regime through the use of FPGAs or ASICs, although further development is needed to reach that stage and current SLM technology would necessitate a reduction in the dynamic range of the output, which is currently 8-bits. The incorporation of the OPP and the automation of the HOC has resulted in a high-speed and robust image recognition system for space situational awareness.

## 6. FUNDING

The work reported here was supported by AFOSR grant No. FA9550-18-01-0359.

## 7. REFERENCES

1. C.-Y. Wang, A. Bochkovskiy, and H.-Y. M. Liao, "YOLOv7: Trainable bag-of-freebies sets new state-of-the-art for real-time object detectors," in *Proceedings of the IEEE/CVF Conference on Computer Vision and Pattern Recognition (CVPR)* (2023), pp. 7464–7475.
2. J. Khoury, M. Cronin-golomb, P. Gianino, and C. Woods, "Photorefractive two-beam-coupling nonlinear joint-transform correlator," J Opt Soc Am B **11**, 2167–2174 (1994).
3. M. S. Monjur, S. Tseng, R. Tripathi, J. J. Donoghue, and M. S. Shahriar, "Hybrid optoelectronic correlator architecture for shift-invariant target recognition," J Opt Soc Am A **31**, 41–47 (2014).
4. M. S. Monjur, S. Tseng, M. F. Fouda, and S. M. Shahriar, "Experimental demonstration of the hybrid opto-electronic correlator for target recognition," Appl. Opt. **56**, 2754–2759 (2017).
5. J. Gamboa, T. Hamidfar, and S. Shahriar, "Integration of a PQ:PMMA holographic memory device into the hybrid opto-electronic correlator for shift, scale, and rotation invariant target recognition," Opt Express **29**, 40194–40204 (2021).
6. J. Gamboa, T. Hamidfar, X. Shen, and S. M. Shahriar, "Elimination of optical phase sensitivity in a shift, scale, and rotation invariant hybrid opto-electronic correlator via off-axis operation," Opt Express **31**, 5990–6002 (2023).
7. D. Casasent and D. Psaltis, "Scale invariant optical correlation using Mellin transforms," Opt Commun **17**, 59–63 (1976).
8. M. S. Monjur, S. Tseng, R. Tripathi, and M. S. Shahriar, "Incorporation of polar Mellin transform in a hybrid optoelectronic correlator for scale and rotation invariant target recognition," J Opt Soc Am A **31**, 1259–1272 (2014).
9. J. Gamboa, M. Fouda, and S. M. Shahriar, "Demonstration of shift, scale, and rotation invariant target recognition using the hybrid opto-electronic correlator," Opt Express **27**, 16507–16520 (2019).